\documentclass[lettersize,journal]{IEEEtran}
\usepackage{amsmath,amsfonts}
\usepackage{algorithmic}
\usepackage{algorithm}
\usepackage{array}
\usepackage[caption=false,font=normalsize,labelfont=sf,textfont=sf]{subfig}
\usepackage{textcomp}
\usepackage{stfloats}
\usepackage{url}
\usepackage{verbatim}
\usepackage{graphicx}
\usepackage{cite}

\begin{document}

\title{DMDC: Dynamic-mask-based dual camera design for snapshot Hyperspectral Imaging}

\author{Zeyu Cai, Chengqian Jin, Feipeng Da
        % <-this % stops a space
\thanks{This work was supported by the National Natural Science Foundation of China under Grants No. 32171911. This work was supported by the National Key R\&D Program of China under Grants 2021YFD2000503. This work was supported by the Special Project on Basic Research of Frontier Leading Technology of Jiangsu Province of China (Grant No. BK20192004C).}% <-this % stops a space
\thanks{Zeyu Cai, Feipeng Da are with Southeast University, Nanjing, china.
E-mails: zeyucai@seu.edu.cn, dafp@seu.edu.cn}
\thanks{Chengqian Jin is with the Chinese Academy of Agricultural Sciences, Nanjing, china.
E-mails: jinchengqian@caas.cn}
}

% The paper headers
\markboth{Journal of \LaTeX\ Class Files,~Vol.~14, No.~8, August~2021}%
{Shell \MakeLowercase{\textit{et al.}}: A Sample Article Using IEEEtran.cls for IEEE Journals}

\IEEEpubid{0000--0000/00\$00.00~\copyright~2021 IEEE}

\maketitle

\begin{abstract}
    Deep learning methods are developing rapidly in coded aperture snapshot spectral imaging (CASSI). The number of parameters and FLOPs of existing state-of-the-art methods (SOTA) continues to increase, but the reconstruction accuracy improves slowly. Current methods still face two problems: 1) The performance of the spatial light modulator (SLM) is not fully developed due to the limitation of fixed Mask coding. 2) The single input limits the network performance. In this paper we present a dynamic-mask-based dual camera system, which consists of an RGB camera and a CASSI system running in parallel. First, the system learns the spatial feature distribution of the scene based on the RGB images, then instructs the SLM to encode each scene, and finally sends both RGB and CASSI images to the network for reconstruction. We further designed the DMDC-net, which consists of two separate networks, a small-scale CNN-based dynamic mask network for dynamic adjustment of the mask and a multimodal reconstruction network for reconstruction using RGB and CASSI measurements. Extensive experiments on multiple datasets show that our method achieves more than 9 dB improvement in PSNR over the SOTA.
   (https://github.com/caizeyu1992/DMDC)
\end{abstract}


%%%%%%%%% BODY TEXT
\section{Introduction}
\label{sec:intro}

\IEEEPARstart{H}{yperspectral} images are applied in astronomy, agriculture, materials, medicine and so on \cite{schechner2021guest, ishida2018novel, wright2019raman, guo2022deep}. Due to the limitation of Nyquist's sampling theorem, the conventional method is not suitable for the spectral reconstruction of dynamic scenes due to the limitation of acquisition time. Inspired by compressive sensing, several compressive spectral imaging techniques have been well developed to improve reconstruction efficiency with unique optical designs and sophisticated processing algorithms. The new methods include snapshot compressive imaging, laminar imaging, and RGB image recovery spectral cubes \cite{cao2016computational}. Among them, coded aperture snapshot spectral imaging (CASSI) is a promising solution.

Existing studies have focused on single-dispersive single-camera CASSI (SD-CASSI) reconstruction algorithms. Although the transformer-based spectral or spatial multi-head self-attention methods are proposed to improve the efficiency of reconstruction. However, because CASSI needs to address the problem of unconstrained optimization algorithms, the single-camera CASSI based on the manual prior mask is difficult to achieve high-quality reconstruction. This problem has received the attention of some researchers and some single-camera improvement systems have been proposed, including colored CASSI \cite{arguello2014colored}, dual-dispersive CASSI \cite{gehm2007single}, and dual camera compressive hyperspectral imaging \cite{chen2022learning} (Fig. \ref{fig:1}). 

Due to the severely ill-posed problem of single-camera, existing algorithms cannot balance accuracy, speed, and flexibility. One solution to this challenge from the hardware side is the dual-camera approach, which captures complementary measurements and mixes this information. However, the existing methods do not take full advantage of the dynamic modulation capability of spatial light modulators (SLM). The SLM used for encoding in the CASSI system is based on LCD or digital micromirror DMD, both LCD and DMD have an 8-bit refresh rate of over 200 HZ and a 1-bit refresh rate of over 20k HZ. Therefore, one of the considerations in this paper is how to exploit the high dynamic capability of SLM and improve the reconstruction capability of the CASSI.

On the other hand, the existing reconstruction algorithm of dual-camera CASSI has not been effectively studied, and the reconstruction effect is comparable to SOTA's SD-CASSI. Firstly, the noise estimation of different sensor acquisitions has not been considered. In addition, the fusion method of RGB and CASSI measurements has yet to be explored.

In this paper, in order to learn the spatial feature of the scene and guide the SLM masking to enhance the capture capability of CASSI. We combine CASSI and an RGB camera to design a dynamic-mask-based dual-camera CASSI (DMDC-CASSI) system. The RGB camera is introduced to both guide the encoding of the SLM and to provide measurement information of other modalities to further aid feature learning and training during reconstruction. The dynamic mask and multimodal information also significantly improve the fidelity of the system. In addition, the asynchronous parallel structure of the RGB optical path and CASSI optical path maintains the snapshot advantage of CASSI.

\IEEEpubidadjcol
Based on the DMDC-CASSI, we propose a multimodal network based on an optical path reversible prior to the reconstruction network design. By the nature of the reversible optical path, we convert the learning of violent mapping into residual learning, and the iterative architecture enhances the learning ability of the dual-stream network. We also designed a dual-stream module consisting of spectral multi-head self-attention, spatial multi-head self-attention, and spectral-spatial multi-head cross-attention by exploiting the local similarity and global correlation of spectra in the spectral and spatial domains. It fully complements the advantages of CASSI, which contains a large amount of encoding information in the spectral dimension, and RGB pictures, which contain a large amount of spatial structure information in the spatial dimension.

The mian contributions of this work can be summarized as follows:

1) We propose a new CASSI system, DMDC-CASSI, using a RGB camera and a CASSI camera in parallel to dynamically adjust the SLM to maximise the loss of critical information during CASSI modulation without changing the CASSI sampling rate.

2) We fuse a reversible prior with a multimodal cross-attention method. By embedding multimodal networks into the reversible prior framework. The RGB images are used to guide CASSI's spectral features to fuse spatial features to improve the recovery of texture information. In addition, the model is based on the aligned CASSI and RGB measurements without prior knowledge of the camera's spectral response, making the network flexible.

3) We evaluate the proposed method on several datasets, including a new publicly available dataset (ARAD\_1K), and we re-evaluated the SOTA model on this dataset. Besides, we provide the RGB dataset for CAVE and KAIST in the TSA-net setting, which is not yet publicly available.

4) Our method achieves a large improvement in the number of parameters and operations compared to previous SOTA methods in different methods. We also evaluate the real-time performance of different networks, and our method also achieves good metrics.
%-------------------------------------------------------------------------
\section{Related Works}
\label{sec:related work}
\subsection{Related works of CASSI system}
\begin{figure}
  \centering
   \includegraphics[width=1\linewidth]{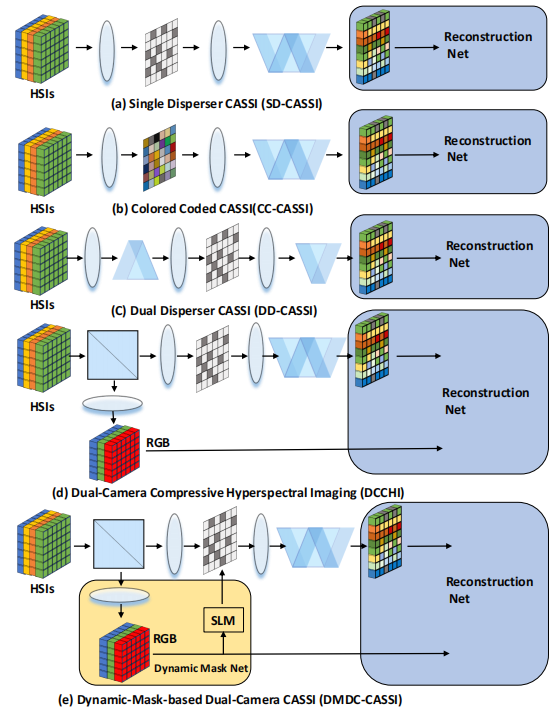}
   \caption{Network framework under different CASSI types.}
   \label{fig:1}
\end{figure}

The typical structure of CASSI is shown in Fig. \ref{fig:2}, where the mask is designed manually and then the spectral 3D cube is used to obtain the measurements through the dispersion device. Single camera CASSI systems are a current research hotspot due to their simple structure, and existing algorithms are largely based on single camera CASSI systems. Based on the single camera CASSI system, Arguello et al. have transformed the encoding of the greyscale into a colored-mask to improve the reconstruction accuracy by additional spectral encoding \cite{arguello2014colored}. Gehm et al. \cite{gehm2007single} used a double dispersioner to spatially encode the spectral channels after the first dispersion and transform the measurements to the original space by inverse secondary dispersion, this spatial-spectral encoding improves the reconstruction quality.

Due to the limitation of the hardware of the single-camera CASSI system, it has been difficult to continue to improve the reconstruction quality of the system by relying solely on algorithms. A part of the study proposed the possibility of joint mask optimization and image reconstruction in order to further improve the reconstruction quality. For instance, Zhang et al. \cite{zhang2020optimization} proposed a constrained optimization-inspired network for adaptive sampling and recovery. HerosNet proposes a network based on learning binary optimal masks, which selects an optimal mask based on the distribution of the dataset to improve the performance of CASSI \cite{zhang2022herosnet}. Another part of the study designed a dual-camera CASSI system, Dual-camera without changing the time cost of single-shot and hardware acquisition, to improve the accuracy and stability of the system by introducing the information of RGB. A more comprehensive introduction about these problems can be found in previous researches \cite{liu2022residual, xie2022dual}. However, both Wang et al. \cite{wang2016adaptive} and He et al. \cite{chen2022learning} simply introduced RGB images into the CASSI system to improve system performance. Besides, the existing dual camera still uses a manual priori mask, which greatly wastes the dynamic modulation capability of the spatial light modulator (SLM). the study of combining dual-camera and asynchronous dynamic mask is still challenging and worth exploring.

\subsection{reconstruction algorithms from CASSI measurement}

CASSI reconstruction algorithms are mostly based on single-camera and include traditional methods \cite{liu2018rank, wang2021tensor, bioucas2007new}, Plug-and-Play (PnP) \cite{7744574, 9495194, Zhang_2019_CVPR, Yuan_2020_CVPR}, end-to-end (E2E) \cite{huang2021deep,meng2020end,cheng2022recurrent,cai2022mask}, and depth unfolding networks (DU) \cite{wang2020dnu, ma2019deep, 9351612, 9879804, meng2020gap}. Traditional methods perform reconstructions based on over-complete dictionaries or sparse spectral features that rely on hand-crafted priors and assumptions. PnP methods combine traditional methods and denoising networks to improve the speed of reconstruction. The E2E are based on CNNs or Transformers and use spectral-spatial local self-similarity and global correlation to achieve a simple and effective reconstruction. DUs use deep-learning modules to simulate traditional convex optimization models and achieve reconstruction through iteration. The above method cannot be directly used for dual camera reconstruction.

In the study of reconstructed methods for dual-camera CASSI. He et al. \cite{he2021fast} use a conventional prior based on the complete dictionary to find spectral basis and spatial coefficients in low-dimensional subspaces, fusing CASSI and RGB measurements. wang et al. \cite{wang2016compressive} used a demosaicing algorithm on a dual-camera design to reconstruct 3D cube. Chen et al. \cite{chen2022learning} transformed the multimodal optimization problem into several simple subproblems by introducing the half quadratic splitting (HQS) alternating optimization method. However, existing algorithms either use the single camera algorithm directly or reconstruct it based on a traditional compression-aware complete dictionary. There exists a poor reconstruction performance and the computational effort is still large.

%-------------------------------------------------------------------------
\section{Methods}
\label{Methods}

\subsection{Model of DMDC-CASSI System}
\label{sec:Model of CASSI System}

\begin{figure*}
  \centering
   \includegraphics[width=0.85\linewidth]{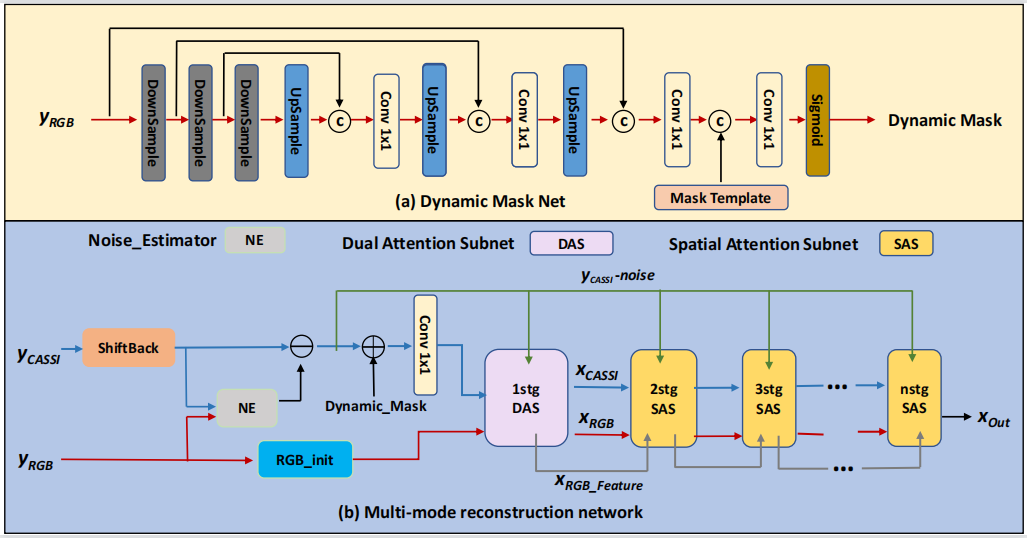}
   \caption{Dynamic Mask Network and Reconstruction Network in DMDC Framework.}
   \label{fig:2}
\end{figure*}

The concise DMDC-CASSI schematic is shown in Fig. \ref{fig:2}(e). Mathematically, considering a 3D HSIs cube, denoted by $X\in \mathbb{R} ^{n_{x}\times n_{y}\times n_{\lambda } }$, where $n_{x}$, $n_{y}$, $n_{\lambda }$ represent the HSIs’s height, width, and number of wavelengths. First, the scene passes through a beamsplitter and by default the energy of the light is distributed equally and without regard to losses, the RGB measurement g(x,y) can be expressed as:

\begin{equation}
 g\left ( x,y \right ) =\frac{1}{2} \int_{\Lambda }\omega \left ( \lambda  \right ) X\left ( x,y,\lambda  \right )d\lambda +  N 
  \label{eq:1}
\end{equation}
where$x\in n^{x},y \in n^{y},\lambda \in n^{\lambda}$, $X\left ( x,y,\lambda  \right )$ is the 3D hyperspectral data cube. $\Lambda$ is the spectral response range of the RGB detector. $\omega \left ( \lambda  \right )$ is the spectral response function of the detector. $N$ is the noise on the RGB detector.

The RGB measurements are passed through a dynamic Mask network that learns the spatial features of the scene and generates a dynamic Mask, which can be expressed as:

\begin{equation}
 M\left ( x,y \right ) =DynamicMaskNet\left ( g(x,y) \right ) 
  \label{eq:2}
\end{equation}
Where $M$ denotes the Mask input to the SLM. The other light path of the spectroscopic prism, which is then modulated by the dynamic mask and a dispersive prism, is finally captured by the CASSI detector. We can express it as:

\begin{equation}
 X^{\prime } (x,y,: ) = \frac{1}{2}M(x,y)\odot X(x,y,:)
  \label{eq:3}
\end{equation}
Where $X^{\prime } \in \mathbb{R}^{n^{x}\times n^{y} \times n^{\lambda }}  $ denotes the modulated spectral data cub.

\begin{equation}
 X^{\prime\prime } (x,y,: ) = X^{\prime }(x,y+d_{\lambda},:)
  \label{eq:4}
\end{equation}

Finally, the captured 2D compressed measurement $Y\in \mathbb{R} ^{n_{x}\times \left ( n_{y}+d\left ( n_{\lambda } -1 \right )   \right )  }$ can be obtained by:

\begin{equation}
\begin{aligned}
Y &=\sum_{\lambda=1}^{n^{\lambda } } X^{\prime \prime } \left ( :,:,\lambda \right ) +G \\
&=\frac{1}{2}\sum_{\lambda=1}^{n^{\lambda } } M(x,y+d_{\lambda})\odot X(x,y+d_{\lambda},:) + G
\end{aligned}
  \label{eq:5}
\end{equation}
where $G\in \mathbb{R} ^{n_{x}\times \left ( n_{y}+d\left ( n_{\lambda }-1  \right )   \right )  }$ is the random noise generated by the photon sensing detector during the measurement.

Combining Eq. \ref{eq:1} and Eq. \ref{eq:5} and rewriting them as linear transformations, the imaging model of dual-camera in the matrix form can be written as:

\begin{equation}
\begin{cases}
 Y_{r} = \Phi _{r} X +N_{r} 
\\
Y_{c} = \Phi _{c} X +N_{c}
\end{cases}
  \label{eq:6}
\end{equation}
Where $Y_{r}$ is the measurement of RGB detector. $N_{r}$ is the noise of RGB detector. $\Phi_{r}$ is the sensing matrix  of RGB detector. $Y_{c}$ is the measurement of CASSI detector. $N_{c}$ is the noise of CASSI detector. $\Phi_{c}$ is the sensing matrix  of CASSI detector, which is generally consider it as the shifted mask.

\begin{figure*}
  \centering
   \includegraphics[width=0.85\linewidth]{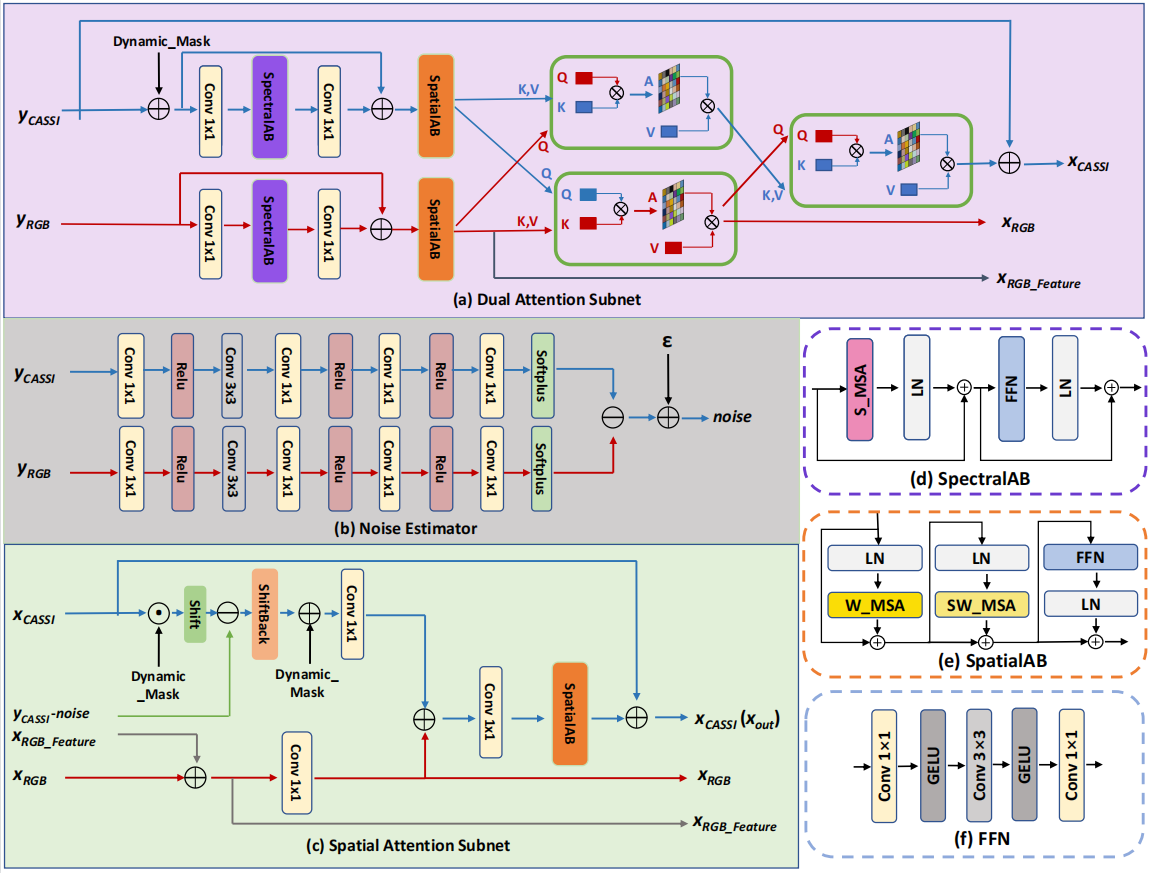}
   \caption{Detailed architecture of our DMDC, including Dual Attention Subnet, Noise Estimator, Spatial Attention SUbnet, SpatialAB, SpectralAB and FFN modules.}
   \label{fig:3}
\end{figure*}

\subsection{Overall architecture of DMDC}

According to the observational model in Eq. \ref{eq:6}, the reconstruction of HSI from DMDC-CASSI and panchromatic images can be considered as an optimisation problem. We recommend a regularization model that can be derived from the observation model:

\begin{equation}
\arg\min \left \| Y_{c}-N_{c} -\Phi _{c}X \right \|_{2}^{2} +  \left \| Y_{r}-N_{r} -\Phi _{r}X \right \|_{2}^{2}
  \label{eq:7}
\end{equation}

According to Eq. \ref{eq:6}, there is a noise in the system for each detector. We assume that the noise in the sensors has the same distribution, which can be estimated using the measurements from both detectors.

\begin{equation}
    \begin{aligned}
noise &=\frac{\Phi_{c}^{-1}Y_{c} -\Phi_{r}^{-1}Y_{r}}{\Phi_{c}^{-1}-\Phi_{r}^{-1}} \\
&=Denoise(Y_{c},Y_{r})
    \end{aligned}
  \label{eq:8}
\end{equation}

Inspired by SST-net \cite{cai2023sst}, considering that the inverse reconstruction process of CASSI is ill-posed, but the forward process of CASSI is well-posed, we reproject the results of each reconstruction back into the measurement space based on the invertible nature of the optical path. The unique solution of the forward process and the actual measurements can create residuals for the network to learn the detailed features of the spectral cube, as described by the following equation:

\begin{equation}
x_{n+1} = f(Y_{c}-N-\Phi _{c}x_{n} ,Y_{r} )+x_{n} 
  \label{eq:9}
\end{equation}

where $x_n$ is the result of the reconstruction after the n-stage iteration. $f$ is an E2E reconstruction module that shares weighted parameters in different phases, except for the initialization phase of the reconstruction.

\subsection{Dynamic mask network}

As shown in Fig. \ref{fig:3}(a), the role of the dynamic mask net is to learn the spatial distribution of the scene from the input RGB image and predict whether each pixel is redundant information and the probability that the pixel can be removed. Also considers the ease of the dynamic mask net. We design a CNN-based network based on the mask template as a framework and finally predict the probability of whether a pixel belongs to the maskable redundant information by sigmoid. The dynamic mask network can be described as follows:

\begin{equation}
\begin{aligned}
M(x,y) & =Sigmoid(Conv(U(y_{rgb}(x,y)\\
& +MaskTemplate(x,y) )))
\end{aligned}
  \label{eq:10}
\end{equation}

\subsection{Multimodal reconstruction network}
The function of the multimodal reconstruction network is to reconstruct the spectral 3D cube from the RGB and CASSI measurements. As shown in Fig. \ref{fig:3}(b), the reconstruction network uses a dual-stream architecture and constructs a residual learning network of Eq. \ref{eq:9} based on the reversible nature of the optical path, and reconstitutes the input of n+1 stage using the results of the n-stage to improve the reconstruction quality. The main modules in Subnet are Noise-Estimator module (NE), RGB-init module, Dual Attention Subnet (DAS) and Spatial Attention Subnet (SAS). First, the network predicts the noise distribution of the detector by the NE module and corrects the CASSI measurements. Subsequently, the role of the RGB initialization module is to map RGB to a higher dimensional space, synchronize the number of RGB channels with the number of CASSI channels, and facilitate learning the long-range dependencies of RGB branches in spectral. Finally, the dual-stream data are iterated to learn the full correlation of the spectrum and the full correlation of the space. DMDC-1stg uses a larger scale DSA module to align the high-dimensional space of RGB with the high-dimensional space of CASSI at a lower computational cost, and is used for RGB to guide the reconstruction of CASSI. After 2stg, the main purpose of reconstruction quality is the elimination of texture and artifacts in high frequencies, and we design SAS module to learn the spatial global similarity of the scene to improve the detailed reconstruction of the model. The network is set up with different size models depending on the number of layers of SAS. DMDC-nstg consists of single DAS and n-1 SAS.

\begin{figure*}
  \centering
   \includegraphics[width=0.88\linewidth]{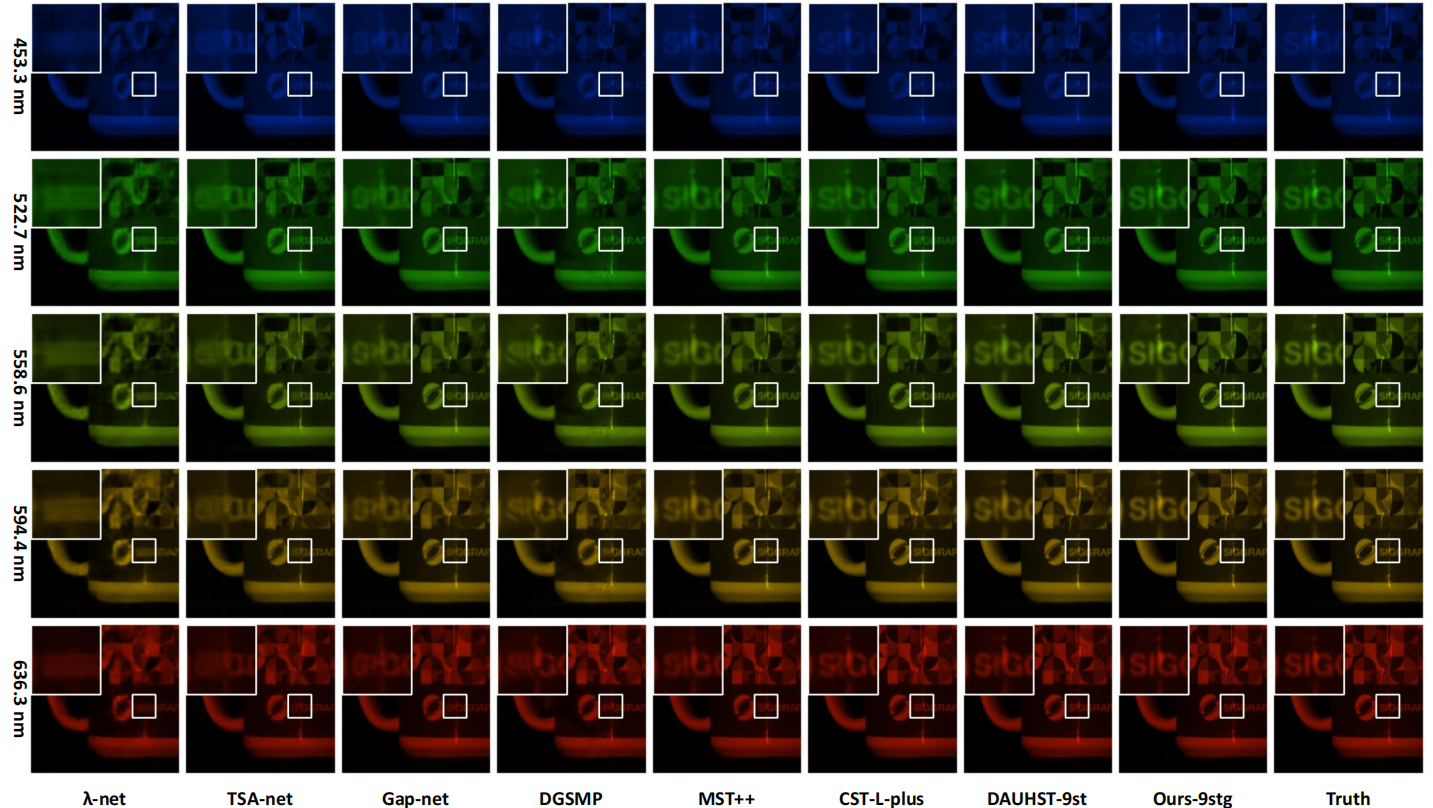}
   \caption{Visual comparisons of DMDC and other SOTA methods of Scene 5 with 5 out of 28 spectral channels on the KAIST dataset. SOTA algorithm from 2019 to 2022 and our DMDCs are included.}
   \label{fig:4}
\end{figure*}

\noindent\textbf{Noise-Estimator module (NE).} 
We assume that the noise on the CASSI detector and the RGB detector have the same distribution function. the NE module maps the CASSI branch and the RGB branch to a high-dimensional space, learns the noise distribution and then downscales it to the original space, and uses the imbalance information between the two branches to estimate the noise. This process can be described as follows:

\begin{equation}
\begin{aligned}
noise &= Soft(Down(Conv(Up(x_{c}))))\\
&-Soft(Down(Conv(Up(x_{r})))) +\epsilon
\end{aligned}
  \label{eq:11}
\end{equation}

\noindent\textbf{RGB-init module.} RGB initialization module, which raises the raw RGB measurements to 64 dimensions and then downscales them to original channels, facilitating the alignment of RGB measurements with CASSI channel counts.

\noindent\textbf{Dual Attention Subnet (DAS).} The DAS module mainly consists of SpectralAB, SpatialAB and CrossAB, because C<<W=H, so spectralAB has the lowest reconstruction cost and is suitable for the initialization of CASSI branches and RGB branches. Based on the spectral features, the global similarity of the two branches in space is learned using spatialAB and CrossAB, respectively, and the features of the two branches are fused by a secondary cross-attention.

\noindent\textbf{Spatial Attention Subnet (SAS).} The residuals of the measurement and the n-stage reconstruction results are constructed, and the global correlation between the residuals and the RGB images in space is learned.

\subsection{Loss function}
Our network has reversible module and reconstruction net, so our loss includes outputting and reversible loss. The outputting loss is calculated as the L2 loss of $x_{out}$ - $x_{truth}$. The reversible loss calculation $x_{out}$ is mapped back to the CCD under the nature of the reversible optical path to obtain the L2 loss of the G ($x_{out}$) value to the actual measurement y. We defined the loss function as follows:

\begin{equation}
\mathcal{L} =\left \| \widehat{x}_{out}- x_{truth}  \right \|_{2}^{2}  + \xi \cdot \left \| \Phi\widehat{x}_{out}-x_{in} \right \|_{2}^{2}  
  \label{eq:12}
\end{equation}
where $\widehat{x} _{out}$ is the predicted values of the network, $\Phi $ represents the process of mask coding and dispersion of predicted values, $x_{in}$ is the measurement of CCD. $\xi$ is the penalty coefficient.

\section{Experiments}

\begin{table*}[]\footnotesize
\begin{center}
\caption{Comparisons between DMDC and SOTA methods on 10 simulation scenes (S1$\sim$S10) in KAIST. PSNR and SSIM are reported.}
\begin{tabular}{ccccccccccccc}
\hline
scene & &S1 & S2  & S3 & S4  & S5  & S6  & S7 & S8  & S9 & S10 & Avg \\
\hline
$\lambda$-net \cite{miao2019net} & PSNR & 32.50  & 31.23 & 33.89 & 40.28 & 29.86 & 30.27 & 30.33 & 28.98 & 31.98 & 28.36 & 31.77 \\
 & SSIM & 89.2\% & 85.4\% & 93.0\%  & 96.5\% & 88.9\% & 89.3\% & 97.5\% & 88.0\%  & 89.1\% & 83.4\% & 89.0\%  \\
  \hline
ADMM-Net \cite{ma2019deep}  & PSNR & 34.12 & 33.62 & 35.04 & 41.15 & 31.82 & 32.54 & 32.42 & 30.74 & 33.75 & 30.68 & 33.58 \\
 & SSIM & 91.8\% & 90.2\% & 93.1\% & 96.6\% & 92.2\% & 92.4\% & 89.6\% & 90.7\% & 91.5\% & 89.5\% & 91.8\% \\
\hline
TSA \cite{meng2020end} & PSNR & 32.95 & 31.69 & 33.01 & 41.24 & 30.12 & 31.89 & 30.75 & 29.89 & 31.61 & 29.9  & 32.30  \\
& SSIM & 91.3\% & 88.4\% & 93.2\% & 97.5\% & 91.1\% & 92.9\% & 89.5\% & 91.2\% & 92.0\%  & 89.0\%  & 91.6\% \\
\hline
Gap-net \cite{meng2020gap} & PSNR & 26.82 & 22.89 & 26.31 & 30.65 & 23.64 & 21.85 & 23.76 & 21.98 & 22.63 & 23.10  & 24.36 \\
 & SSIM & 75.4\% & 61.0\%  & 80.2\% & 85.2\% & 70.3\% & 66.3\% & 68.8\% & 65.5\% & 68.2\% & 58.4\% & 66.9\% \\
\hline
DGSMP \cite{huang2021deep} & PSNR & 33.26 & 32.09 & 33.06 & 40.54 & 28.86 & 33.08 & 30.74 & 31.55 & 31.66 & 31.44 & 32.63 \\
 & SSIM & 91.5\% & 89.8\% & 92.5\% & 96.4\% & 88.2\% & 93.7\% & 88.6\% & 92.3\% & 91.1\% & 92.5\% & 91.7\% \\
  \hline
PnP-DIP-HSI \cite{meng2021self}  & PSNR & 32.70 & 27.27 & 31.32 & 40.79 & 29.81 & 30.41 & 28.18 & 29.45 & 34.55 & 28.52 & 31.30 \\
 & SSIM & 89.8\% & 83.2\% & 92.0\% & 97.0\% & 90.3\% & 89.0\% & 91.3\% & 88.5\% & 93.2\% & 86.3\% & 90.1\% \\
 \hline
BIRNAT \cite{cheng2022recurrent}  & PSNR & 36.79 & 37.89 & 40.61 & 46.94 & 35.42 & 35.30 & 36.58 & 33.96 & 39.47 & 32.80 & 37.58 \\
 & SSIM & 95.1\% & 95.7\% & 97.1\% & 98.5\% & 96.4\% & 95.9\% & 95.5\% & 95.6\% & 97.0\% & 93.8\% & 96.0\% \\
\hline
HDNet \cite{hu2022hdnet}   & PSNR & 35.14 & 35.67 & 36.03 & 42.30  & 32.69 & 34.46 & 33.67 & 32.48 & 34.89 & 32.38 & 34.97 \\
 & SSIM & 93.5\% & 94.0\%  & 94.3\% & 96.9\% & 94.6\% & 95.2\% & 92.6\% & 94.1\% & 94.2\% & 93.7\% & 94.3\% \\
\hline
MST++ \cite{cai2022mst++}     & PSNR & 35.53 & 35.68 & 35.99 & 42.78 & 32.71 & 35.14 & 34.24 & 33.30  & 35.13 & 32.86 & 35.34 \\
 & SSIM & 94.6\% & 94.6\% & 95.4\% & 97.7\% & 94.9\% & 95.9\% & 93.8\% & 95.7\% & 95.1\% & 94.8\% & 95.3\% \\
\hline
CST-L \cite{lin2022coarse}     & PSNR & 35.96 & 36.84 & 38.16 & 42.44 & 33.25 & 35.72 & 34.86 & 34.34 & 36.51 & 33.09 & 36.12 \\
 & SSIM & 94.9\% & 95.5\% & 96.2\% & 97.5\% & 95.5\% & 96.3\% & 94.4\% & 96.1\% & 95.7\% & 94.5\% & 95.7\% \\
\hline
HerosNet \cite{zhang2022herosnet}  & PSNR & 35.69 & 35.01 & 34.82 & 38.07 & 33.18 & 34.94 & 33.58 & 33.19 & 33.04 & 33.01 & 34.45 \\
 & SSIM & 97.3\% & 96.8\% & 96.7\% & 98.5\% & 96.9\% & 97.6\% & 96.2\% & 96.8\% & 96.4\% & 96.5\% & 97.0\% \\
\hline
DAUHST-9stg \cite{cai2022degradation} & PSNR & 37.25 & 39.02 & 41.05 & 46.15 & 35.80  & 37.08 & 37.57 & 35.10  & 40.02 & 34.59 & 38.36 \\
 & SSIM & 95.8\% & 96.7\% & 97.1\% & 98.3\% & 96.9\% & 97.0\%  & 96.3\% & 96.6\% & 97.0\%  & 95.6\% & 96.7\% \\
 \hline
SST-LPlus \cite{cai2023sst}  & PSNR & 38.24 & 40.05 & 42.45 & 47.87 & 37.02 & 37.59 & 37.20 & 35.42 & 40.54 & 35.25 & 39.16 \\
 & SSIM & 96.8\% & 97.4\% & 97.9\% & 99.2\% & 97.5\% & 97.5\% & 96.0\% & 97.1\% & 97.5\% & 96.7\% & 97.4\% \\
\hline
RDLUF-9stage \cite{Dong_2023_CVPR}  & PSNR & 37.94 & 40.95 & 43.25 & 47.83 & 37.11 & 37.47 & 38.58 & 35.50 & 41.83 & 35.23 & 39.57 \\
 & SSIM & 96.6\% & 97.7\% & 97.9\% & 99.0\% & 97.6\% & 97.5\% & 96.7\% & 97.0\% & 97.8\% & 96.2\% & 97.4\% \\
\hline
DMDC-1stg  & PSNR &41.37 & 40.50 & 36.76 & 42.62 &40.10  &43.94  & 40.39 & 40.05 & 37.99 & 40.92 & 40.47 \\
 & SSIM & 99.0\% & 99.1\% & 97.8\% & 99.1\%  & 99.4\% & 99.6\% & 98.5\% & 99.2\% & 98.4\% & 99.6\% & 99.0\% \\

DMDC-3stg  & PSNR & 45.44 & 47.31 & 44.42 & 48.66& 43.75 & 47.38 & 44.78 & 43.58& 41.51 & 45.63 & 45.25 \\
 & SSIM & 99.6\% & 99.7\% & 98.8\% & 99.7\%  & 99.6\% & 99.8\% & 99.1\% & 99.6\% & 99.0\% & 99.8\% & 99.5\% \\

DMDC-5stg  & PSNR & 48.31 & 53.01 & 47.26 & 54.57& 46.78 & 48.48 & 46.04 & 46.87& 47.81 & 48.89 & 48.80 \\
 & SSIM & \pmb{99.7\%} & \pmb{99.9\%} & 99.0\% & \pmb{99.9\%}  & \pmb{99.8\%} & 99.8\%& \pmb{99.2\%} & 99.7\% & \pmb{99.5\%} & \pmb{99.9\%} & 0.996  \\

DMDC-7stg  & PSNR & 48.47 & 53.45 & 47.55 &\pmb{54.62} & \pmb{47.31} & 48.60 & 45.94 &47.31 & 47.66 & 49.08 & 49.00 \\
 & SSIM &\pmb{ 99.7\%} & \pmb{99.9\%} & \pmb{99.1\%} &  \pmb{99.9\%} & \pmb{99.8\%} &99.8\% & \pmb{99.2\%}& \pmb{99.8\%} &\pmb{99.5\%}  &\pmb{99.9\%} & \pmb{99.7\%}  \\

DMDC-9stg  & PSNR &\pmb{ 48.51} & \pmb{53.67} & \pmb{47.68} & 54.42& 47.22 & \pmb{48.70} & \pmb{46.22} & \pmb{47.39}& \pmb{48.38} & \pmb{49.19} & \pmb{49.14} \\
 & SSIM &\pmb{ 99.7\%} & \pmb{99.9\%} & \pmb{99.1\%} & \pmb{99.9\%}  & \pmb{99.8\%} & \pmb{99.9\%} & \pmb{99.2\%} & \pmb{99.8\%} & \pmb{99.5\%} & \pmb{99.9\%} & \pmb{99.7\% } \\

\hline
\label{tab:1}
\end{tabular}
\end{center}
\end{table*}

\begin{figure}[t]
  \centering
 \includegraphics[width=1\linewidth]{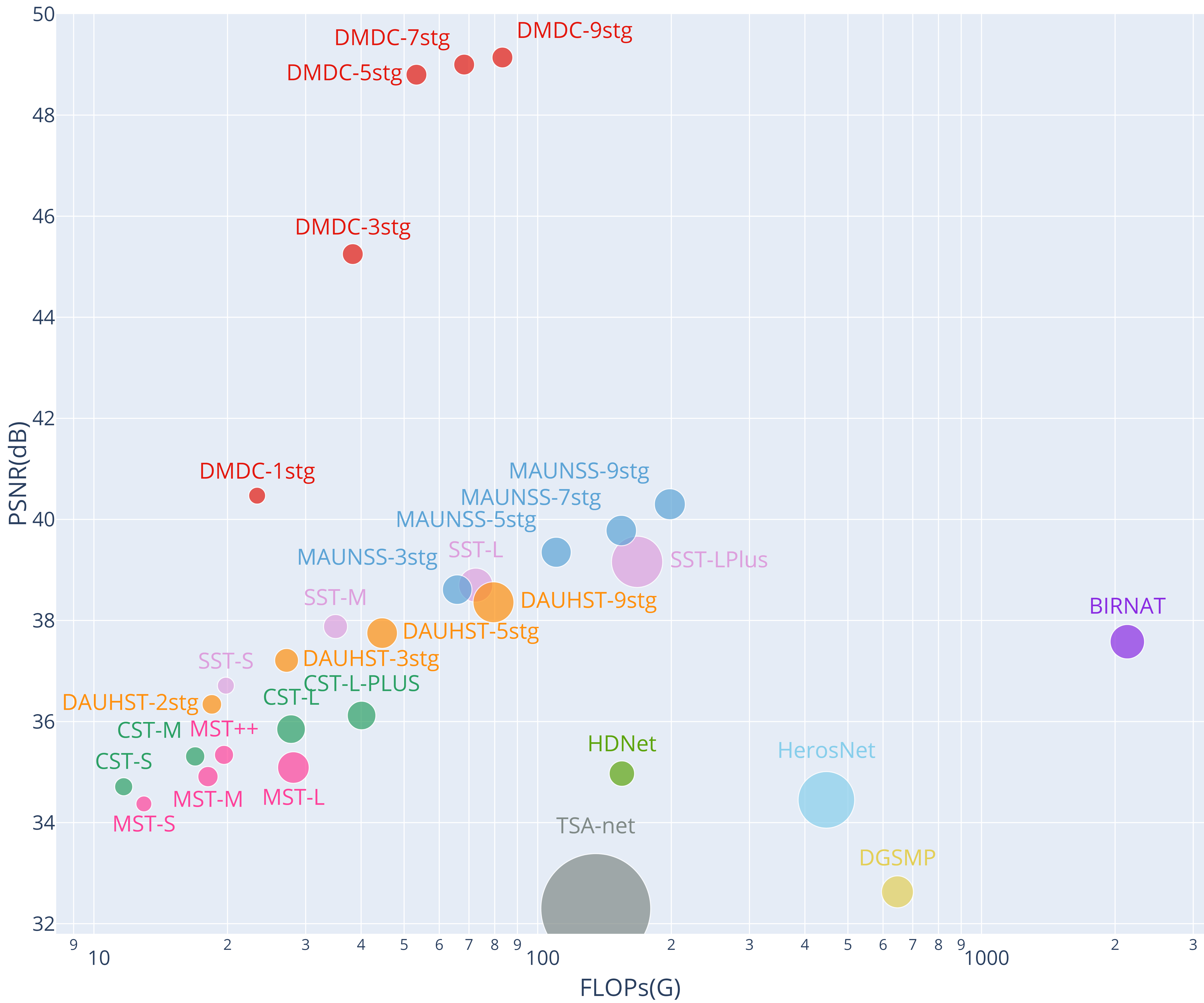}
   \caption{PSNR-Params-FLOPs comparisons of our DMDCs and SOTA HSI reconstruction methods. The vertical axis is PSNR (dB), the horizontal axis is FLOPs (computational cost), and the circle radius is Params (memory cost).}
   \label{fig:0}
\end{figure}

\begin{figure}
  \centering
   \includegraphics[width=1\linewidth]{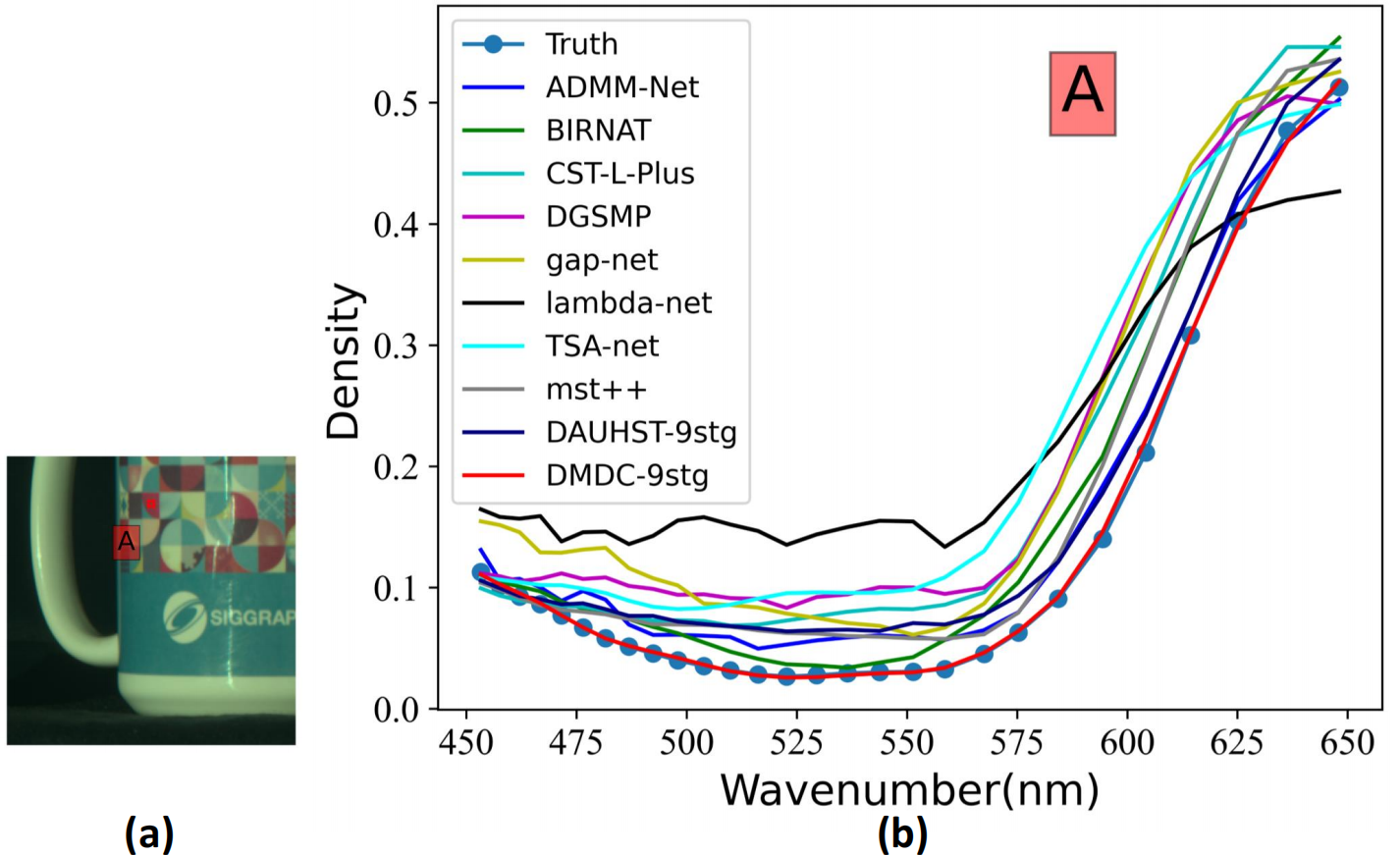}
   \caption{Spectral curves on the selected regions A and the visualization results show that our results have higher spectral accuracy and better perceptual quality.}
   \label{fig:5}
\end{figure}

\begin{figure*}
  \centering
   \includegraphics[width=0.88\linewidth]{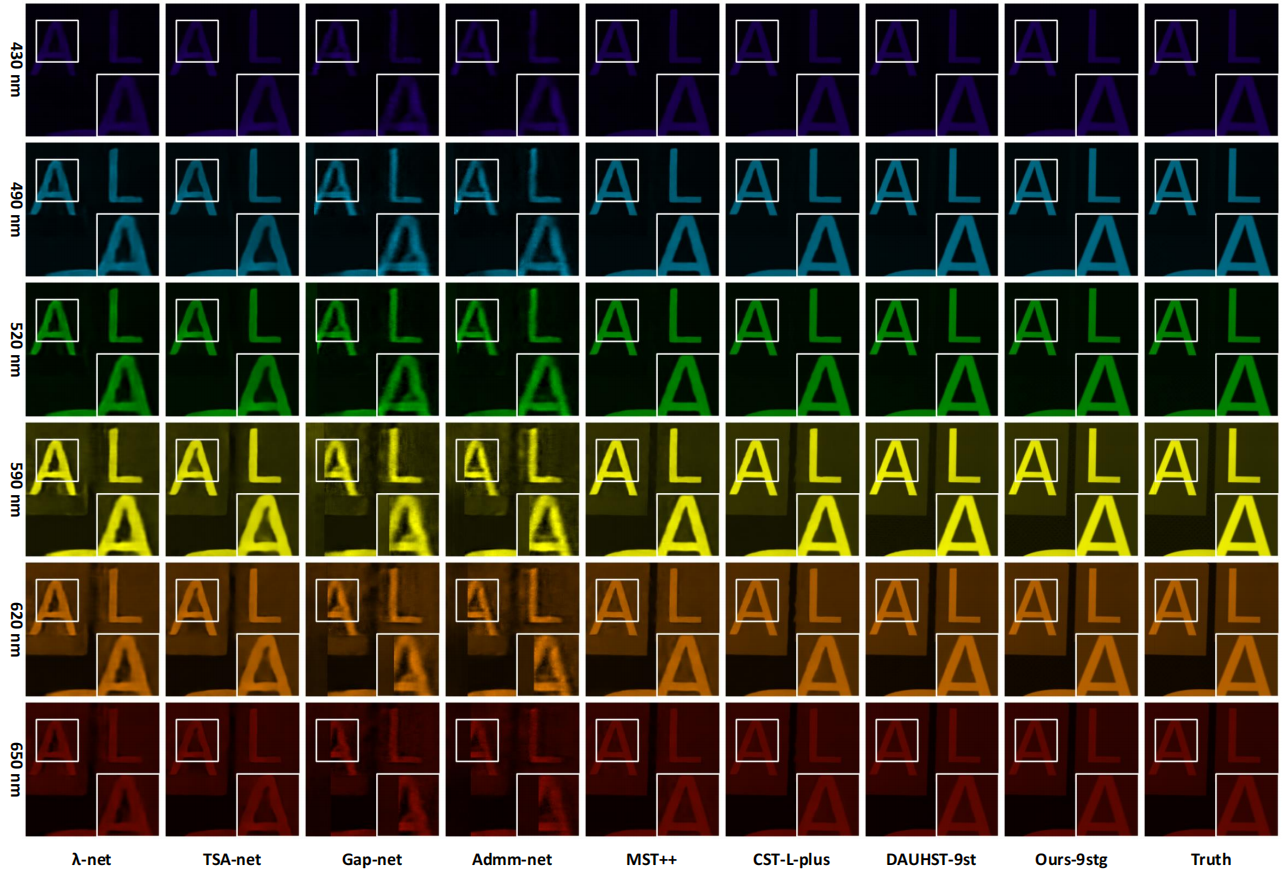}
   \caption{Visual comparisons of our DMDC and other SOTA methods of Scene 33 with 6 out of 31 spectral channels on the ARAD\_1K dataset. SOTA algorithm and our DMDCs are reported.}
   \label{fig:6}
\end{figure*}

In this section, we experiment on multiple datasets to demonstrate the advantages of the proposed method over other SOTA methods in reconstruction. The RGB data used in CAVE and KAIST is synthesized, and the data will be exposed in the code. In addition, we re-evaluated the SOTA model in CASSI on the latest dataset ARAD\_1K, which is the first large-scale evaluation of the CASSI algorithm since the release of this dataset.

\subsection{Experiment Setup}

\noindent\textbf{Simulated Experiments On CAVE and KAIST.} In the first testing dateset, there are 28 spectral channels, with wavelengths ranging from 450 nm to 650 nm. We use two simulation hyperspectral image datasets, CAVE\cite{park2007multispectral} and KAIST\cite{choi2017high}. The CAVE dataset consists of 32 hyperspectral pictures with 512 $\times$ 512 spatial resolution. The KAIST dataset consists of 30 hyperspectral pictures with a 2704 $\times$ 3376 pixel spatial resolution. We use CAVE as the training set while adhering to the TSA-Net schedule. For testing, 10 scenes from KAIST were chosen.

\noindent\textbf{Experiments On ARAD\_1K.} A larger-than-ever natural hyperspectral image data set is offered in the dataset known as ARAD\_1K \cite{arad2022ntire}. ARAD\_1K is made up of 31 hyperspectral pictures with a spatial resolution of 482 $\times$ 512 and a 10 nm step from 400 to 700 nm. The data set's 1,000 measurements were separated into the following categories: 50 test measurements and 900 training measurements.

\noindent\textbf{Evaluation Metrics.}  We adopt peak signal-to-noise ratio (PSNR), structural similarity (SSIM), Mean Relative Absolute Error (MRAE) and Root Mean Square Error (RMSE) as the metrics to evaluate the HSI reconstruction performance.
As in previous works, the main evaluation metrics of CAVE and KAIST are PSNR, SSIM, parameters and floating point numbers. The main evaluation metric of ARAD\_1K is MRAE, PSNR and other metrics are used as reference. We also introduce Frames Per Second (FPS) as a reference in ARAD\_1K.

\noindent\textbf{Implementation Details.}  We implement DMDC-net in Pytorch. Our DMDC-1stg, DMDC-3stg, DMDC-5stg, DMDC-7stg, DMDC-9stg and other SOTA mrthods are trained on 1 $\times$ RTX 3090 GPU. We adopt Adam optimizer ($\beta_{1} = 0.9$ and $\beta_{2} = 0.999$) for 300 epochs. The learning rate is set to $4\times 10^{-4}$ in the beginning and is halved every 50 epochs during the training procedure.

\begin{table*}[]\footnotesize
\begin{center}
\caption{Comparisons between DMDC-net and SOTA methods on 50 scenes in ARAD\_1K. MRAE, RMSE, PSNR, SSIM , Params, FLOPS and reconstruction time are reported.}
\begin{tabular}{ccccccccccccc}
\hline
method  & Avg MRAE & Avg RMSE     & Avg PSNR     & Avg SSIM     & Params(M)     & GFLOPS(G)     & Time for 50 measurements (s)\\
\hline
$\lambda$-net \cite{miao2019net} & 0.1664 & 0.0228 & 33.29 & 91.63\% & 32.73 & 31.18 & 0.19  \\
\hline
TSA \cite{meng2020end}  & 0.1392 & 0.0193 & 34.82 & 93.82\% & 44.30 & 142.34 & 0.53 \\

\hline
Gap-net \cite{meng2020gap} & 0.3005 & 0.0466 & 27.02 & 84.26\% & 4.28 & 87.55 & 1.10 \\
\hline
Admm-net \cite{ma2019deep}  & 0.2949 & 0.0475 & 27.03 & 84.22\% & 4.28 & 87.55 & 1.02 \\

\hline
MST-S \cite{cai2022mask}  & 0.0984 & 0.0135 & 37.93 & 96.51\% & 1.12 & 14.93 & 0.77 \\

MST-M \cite{cai2022mask}  & 0.0855 & 0.0119 & 39.02 & 97.21\%  & 1.81 & 20.85 & 1.33 \\

MST-L \cite{cai2022mask} & 0.0776 & 0.0102 & 40.64 & 97.71\% & 2.45 & 32.18 & 1.97 \\

MST++ \cite{cai2022mst++}  & 0.0936 & 0.0127 & 38.49 & 96.92\% & 1.63 & 23.88 & 0.55 \\

\hline
CST-S \cite{lin2022coarse} & 0.1308 & 0.0165 & 36.25 & 95.23\% & 1.46 & 12.21 & 0.64 \\

CST-M \cite{lin2022coarse}  & 0.1147 & 0.0153 & 36.91 & 95.77\% & 1.66 & 15.96 & 1.17 \\

CST-L \cite{lin2022coarse}  & 0.0956 & 0.0125 & 38.95 & 96.74\% & 3.66 & 27.17 & 1.68 \\

CST-LPlus \cite{lin2022coarse}  & 0.0854 & 0.0115 & 39.63 & 97.16\% & 3.66 & 33.95 & 2.15 \\

\hline
DAUHST-3stg \cite{cai2022degradation}& 0.0686 & 0.0094 & 41.31 & 97.90\% & 1.64 & 29.13 & 0.99 \\

DAUHST-5stg \cite{cai2022degradation} & 0.0590 & 0.0076 & 43.89 & 98.36\% & 2.70 & 47.72 & 1.50 \\

DAUHST-7stg \cite{cai2022degradation}& 0.0575 & 0.0074 & 44.20 & 98.44\% & 3.77 & 66.32 & 3.62 \\

DAUHST-9stg \cite{cai2022degradation}& 0.0602 & 0.0073 & 44.18 & 98.45\% & 4.83 & 84.92 & 4.42 \\
\hline
DMDC-1stg  & 0.0747 & 0.0093 & 41.36 & 98.63\% & 1.26 & 27.11 & 0.54\\

DMDC-3stg  & 0.0399 & 0.0043 & 48.12 & 99.39\% & 1.92 & 45.45 & 2.57 \\

DMDC-5stg & 0.0360 & \pmb{0.0038} & 49.11 & \pmb{99.49\%} & 1.92 & 63.78 & 3.91 \\

DMDC-7stg & 0.0361 & \pmb{0.0038} & 49.08 & \pmb{99.49\%} & 1.92 & 82.11 & 5.21 \\

DMDC-9stg & \pmb{0.0357} & \pmb{0.0038} & \pmb{49.14} & \pmb{99.49\%} & 1.92 & 100.44 & 6.53 \\

\hline
\label{tab:1}
\end{tabular}
\end{center}
\end{table*}

\begin{figure*}
  \centering
   \includegraphics[width=0.8\linewidth]{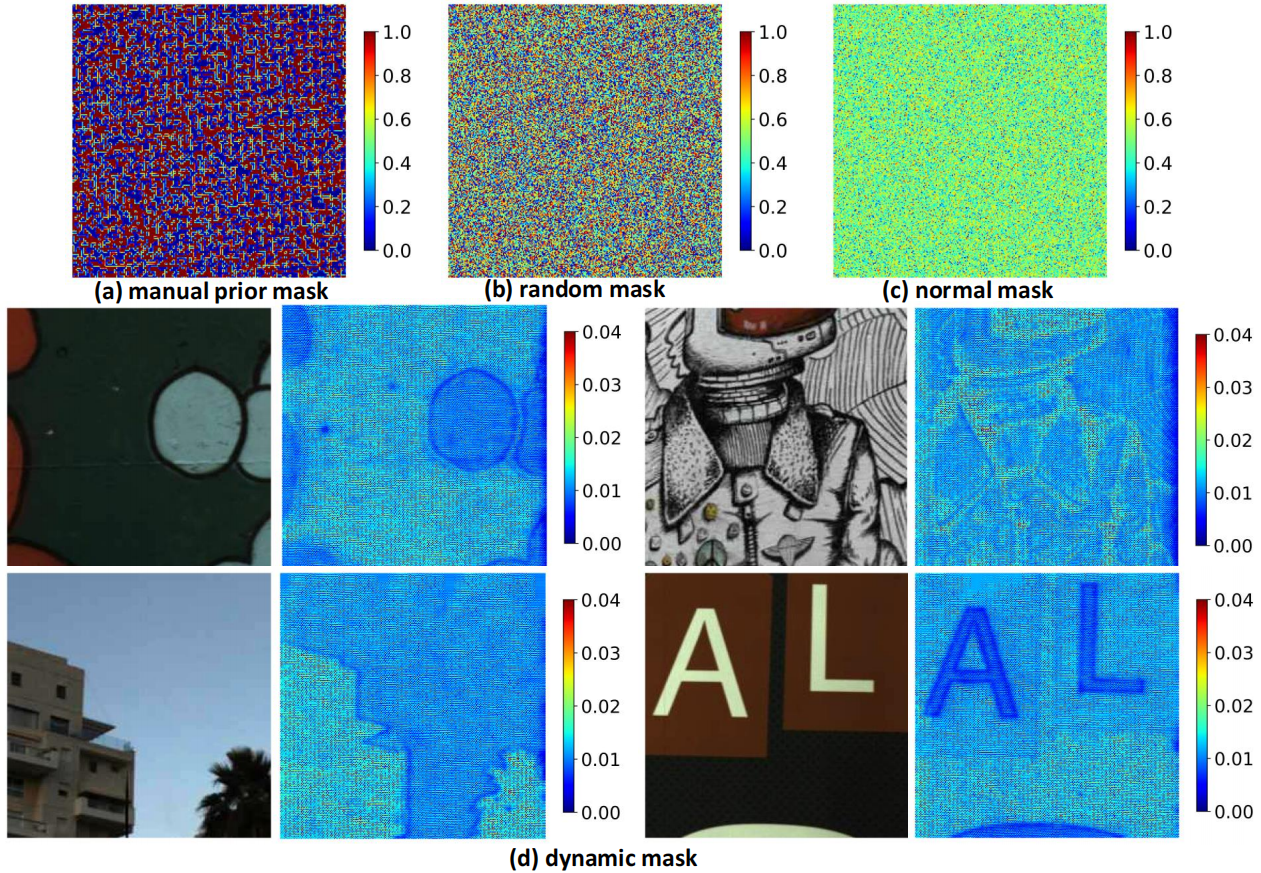}
   \caption{Comparison of manual prior mask, random mask, normal mask and dynamic mask.}
   \label{fig:7}
\end{figure*}

\begin{figure}
  \centering
 \includegraphics[width=1\linewidth]{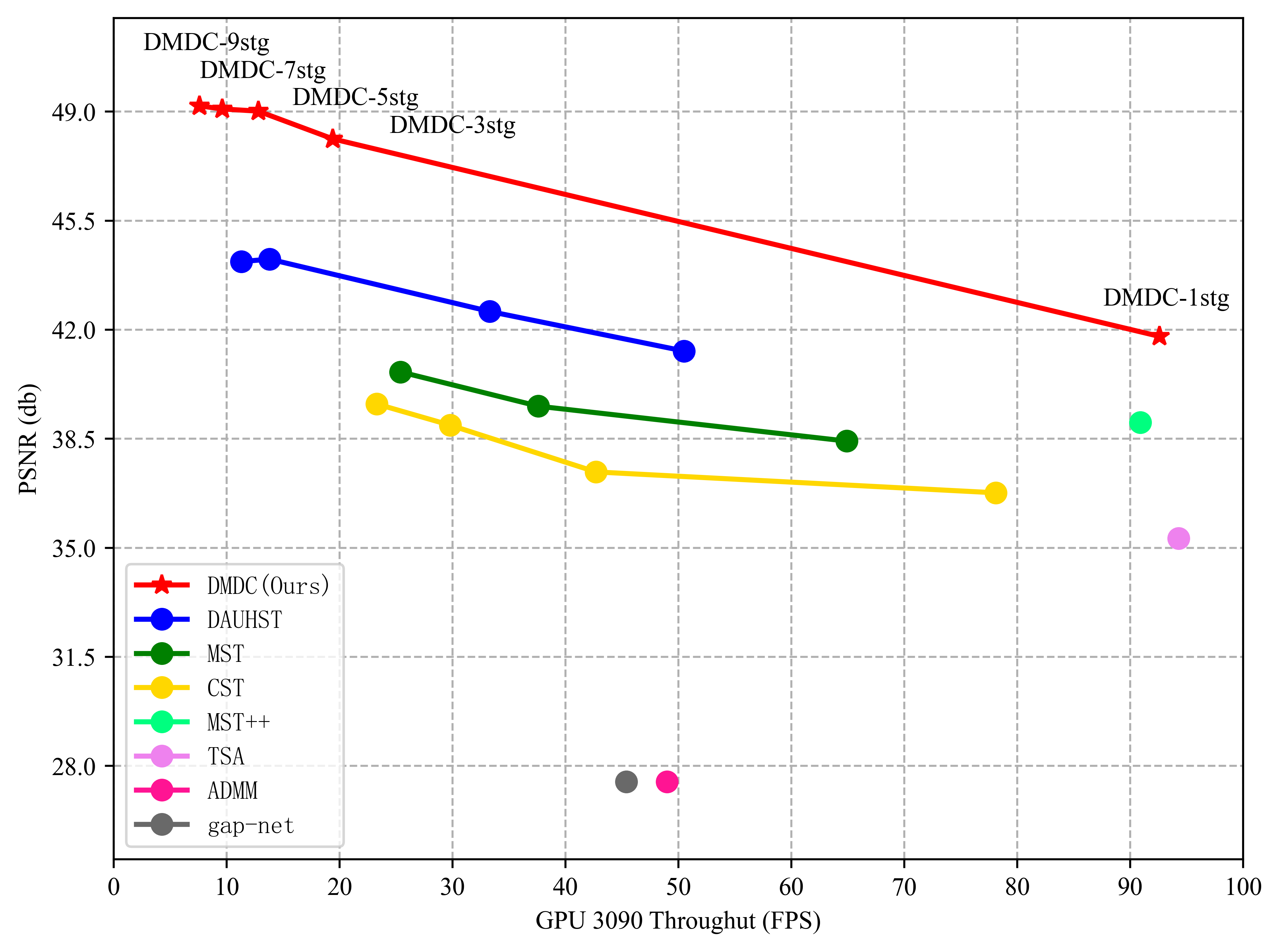}
   \caption{Comparison of DMDC and other SOTA spectral reconstructed methods. Both FPS and PSNR are given for a handy reference. All models are test with 3090 on ARAD\_1K.}
   \label{fig:8}
\end{figure}

\begin{table*}[h]\footnotesize
  \begin{center}
  \caption{Evaluation of the effectiveness of RGB branch and Noise Esitimator.}
  \begin{tabular}{lccccccccc}
    \hline
    Base-line & Dynamic Mask & RGB & Noise Estimator & MRAE & RMSE & PSNR & SSIM & Params & GFLOPs  \\
     & & & & Avg & Avg & Avg(dB) & Avg & (M) &  \\
    \hline
    DMDC-3stg & $\surd$ & $\surd$ & $\times$ & 0.0437 & 0.0050 & 46.65 & 99.33\%  & 1.83 & 39.10 \\
    DMDC-3stg & $\surd$ & $\times$ & $\surd$  & 0.0685 & 0.0071 & 43.52 & 98.91\% & 1.87 & 42.18 \\
    DMDC-3stg & $\surd$ & $\surd$ & $\surd$  & 0.0398 & 0.0045 & 47.45 & 99.42\%  & 1.92 & 45.45 \\
    \hline
  \end{tabular}
  \label{tab:3}
\end{center}
\end{table*}

\begin{table*}[h]\footnotesize
    \begin{center}
    \caption{Evaluation of the effectiveness of Cross-Attention in different sizes of DMDC.}
  \begin{tabular}{lccccccc}
    \hline
    Base-line & With Cross-Attention & Avg MRAE & Avg RMSE & Avg PSNR & Avg SSIM & Params & GFLOPs  \\
     & &  &  & (dB) &  & (M) &  \\
    \hline
    DMDC-1stg & $\surd$ &  0.0747 & 0.0093 & 41.36 & 98.63\%  & 1.26 & 27.11 \\
    DMDC-1stg & $\times$ & 0.0839 & 0.0096 & 40.86 & 98.44\% & 1.24 & 24.95 \\
    DMDC-3stg & $\surd$ & 0.0399 & 0.0043 & 48.12 & 99.39\%  & 1.92 & 45.45 \\
    DMDC-3stg & $\times$ & 0.0416 & 0.0046 & 47.38 & 99.38\%  & 1.90 & 43.28 \\
    \hline
  \end{tabular}
  \label{tab:4}
\end{center}
\end{table*}

\begin{table*}[h]\footnotesize
\begin{center}
  \caption{Evaluation of the effectiveness of different types of Mask.}
  \begin{tabular}{lccccccccc}
    \hline
    Base-line & Mask type & Avg MRAE & Avg RMSE & Avg PSNR & Avg SSIM & Params & GFLOPs \\
     & &  &  & (dB) & & &  \\
    \hline
    DMDC-3stg & Manual Prior Mask &  0.0567 & 0.0066 & 44.12 & 98.97\% & 1.92 & 45.45 \\
    DMDC-3stg & Rand Mask & 0.0596 & 0.0062 & 44.94 & 98.86\% & 1.92 & 45.45 \\
    DMDC-3stg & Normal Mask & 0.0589 & 0.0070 & 43.93 & 98.85\% & 1.92 & 45.45 \\
    DMDC-3stg & Dynamic Mask & 0.0399 & 0.0043 & 48.12 & 99.39\% & 1.92 & 45.45 \\
    \hline
  \end{tabular}
  \label{tab:5}
\end{center}
\end{table*}

\subsection{results on CAVE and KAIST}
\textbf{(i)}On CAVE and KAIST, our best model DMDC-9stg yields very impressive results, i.e., 49.14 dB in PSNR and 99.7\% in SSIM ,and it is the only one with the PSNR of more than 49 dB, which is more than 9 dB than the best PSNR of the SOTA published models, and the SSIM is more than 2.3\%. DMDC-9stg significantly outperforms RDLUF-9stage, SST-LPlus, DAUHST-9stg, HerosNet, CST-L, MST++, DNet, BIRNAT, DIP-HSI, DGSMP, GAp-net, TSA-net, ADMM-Net, and $\lambda$-net of PSNR by 9.57, 9.98, 10.78, 14.69, 13.02, 13.80, 14.17, 11.56, 17.88, 16.51, 24.78, 16.84, 15.56 and 17.37 dB, and 2.3\%, 2.3\%, 3.0\%, 2.7\%, 4.0\%, 4.4\%, 5.4\%, 3.7\%, 10.3\%, 8.0\% , 32.8\% , 8.1\% , 7.9\% , and 10.7\% improvement of SSIM. Detailed comparison data are shown in Tab. \ref{tab:1}.

\textbf{(ii)}It can be observed that our DMDC-net significantly surpass SOTA methods by a large margin while requiring much cheaper memory and computational costs. Compared with DU methods, DMDC-5stg outperforms RDLUF-9stage by 9.23 db but only costs 85.9\% (1.59/1.85) params and 44.4\% (53.30/120.08) FLOPs, outperforms DAUHST-9stg 10.44 db but only costs 25.8\% (1.59/6.15) params and 67.0\% (53.30/79.50) FLOPs. And outperforms other Transformer-based methods, our DMDC-1stg outperforms CST-L by 4.62 dB but only costs 35.0\% (1.05/3.00) Params and 83.8\% (23.32/27.81) FLOPs, and DMDC-1stg outperforms MST++ by 5.13 dB, but only costs 28.7\% (1.05/3.66) Params and 82.8\% (23.32/28.15) FLOPs. In addition, our DMDC-3stg, DMDC-5stg, DMDC-7stg and DMDC-9stg outperforms other competitors by very large margins. we provide PSNR-Params-FLOPs comparisons of different reconstruction algorithms in Fig. \ref{fig:1}.

 In addition, as illustrated in Fig. \ref{fig:5}(a), in A positions, although all the restoration algorithms can better describe the qualitative trend of spectral changes, the spectral curves of the DMDC have a higher correlation with the reference spectra in the quantitative spectral data. As shown in Fig. \ref{fig:5}(b) with the red line and the blue line.

\subsection{Results on ARAD\_1K}
On ARAD\_1K, the comparison data of DMDC and SOTA models are shown in Table 2, our DMDC-9stg yields 0.0357 in MRAE, 0.0038 in RMSE, 49.14 in PSNR, 99.49\% in SSIM. Due to the larger size of ARAD\_1K compared to CAVE and KAIST, the reconstruction quality of many models has been improved. DMDC-9stg still significantly outperforms DAUHST-9stg, CST-LPlus, MST++, Admm-net, Gap-net, TSA-net,  and $\lambda$-net of MRAE by 0.0245, 0.0497, 0.0579, 0.2592, 0.2648, 0.1035, and  0.1307, and of PSNR by  4.96, 9.51, 10.65, 22.11, 22.12, 14.32, and 15.85, and 1.04\%, 2.33\%, 2.57\%, 15.27\%, 15.23\%, 5.67\%, and 7.86\% improvement of SSIM, suggesting the effectiveness of our method.
Fig. \ref{fig:6} plots the visual comparisons of our DMDC-9stg and other SOTA methods on Scene 32 with 6 (out of 31) spectral channels. The bottom-left part shows the zoomed-in patches of the white boxes in the entire HSIs, the reconstructed HSIs produced by DMDCs have more spatial details and clearer texture in different spectral channels than other SOTA methods.  In particular, the inner edges of A in the figure have some curvature except for DMDCs.

\subsection{Ablation study}
To evaluate the contribution of different components in the proposed DMDCs, ablation study is conducted on the ARAD\_1K datasets. We mainly focus on the four components, namely Dynamic mask, RGB branching, Noise Estimator and Cross Attention. Tab. \ref{tab:3} shows the comparsion results of the proposed networks DMDC-3stg on different setting of RGB branching and Noise Estimator. From Tab. \ref{tab:3}, we can observe that using the Noise Estimator can improve the reconstruction result by 0.80 db. And When using RGB, RMAE can be reduced by half, while PSNR can be increased by 3.93 db, SSIM similarity is increased by 0.51\%. 

Tab. \ref{tab:4} shows the importance of Cross-Attention at different sizes of DMDC, providing MRAE, RMSE, PSNR and SSIM.  From Tab. \ref{tab:4}, we can observe that Cross-Attention can improve the reconstruction result of DMDC-1stg and DMDC-3stg by 1.00 db and 0.74 db, respectively.

Tab. \ref{tab:5} shows the results of MRAE, RMSE, PSNR and SSIM on different Mask type of DMDC, including manual prior mask, random mask, normal mask and dynamic mask. The results show that the dynamic mask contributes about as much to the network boost as the RGB branching. The difference between dynamic mask and other fixed masks for different scenes is shown in Fig. \ref{fig:7}. dynamic mask has a better encoding method based on RGB prior.

To evaluate the combined comparison of the reconstruction speed and performance of different algorithms on  ARAD\_1K, we analyzed the PSNR and FPS of SOTA methods. As shown in Fig. \ref{fig:8}, our DMDC-1stg is twice as fast for the same reconstruction quality. In addition, our DMDC-3stg, DMDC-5stg, DMDC-7stg, and DMDC-9stg have higher reconstruction accuracy at the same reconstruction frame rate. Especially the DMDC-5stg, with the PSNR of over 49 db, and the frame rate is over 12.

\section{Conclusion}
In this paper, we propose a new CASSI framework, DMDC. the new framework mainly addresses the two problems of insufficient dynamic capability of SLM and multimodal fusion of CASSI measurements with RGB measurements. Based on the DMDC framework, we propose a dynamic mask network and a Multimodal reconstruction network, and in order to fuse CASSI measurements with RGB images, we embed the Dual In order to fuse CASSI measurements with RGB images, we embed Dual Attention subnet, Noise Estimator and Spatial Attention subnet in Multimodal reconstruction network. With these novel techniques, we establish a series of extremely efficient DMDC-Net models. Quantitative experiments demonstrate that our method surpasses SOTA algorithms by a large margin, even requiring significantly cheaper Params and FLOPs. DMDC demonstrates that multimodality can solve the CASSI ill-posed problem in one-shot imaging, and future work will focus on reducing the cost of CASSI and increasing the speed of reconstruction.

%\section*{Acknowledgments}

\bibliographystyle{IEEEtran}
\bibliography{IEEEabrv,mybibfile.bib}

\vfill

\end{document}